\documentclass[11pt]{article}
\usepackage{epsfig,psfig,float,amssymb,latexsym}
\usepackage[notref,notcite]{}
\input{epsf}

\newcommand{\mpi}{\mu}
\newcommand{\NP}[1]{ Nucl.\ Phys.\ {\bf #1}}

\newcommand{\PL}[1]{ Phys.\ Lett.\ {\bf #1}}

\newcommand{\AN}[1]{Ann. Phys. NY {\bf #1}}

\newcommand{\PR}[1]{Phys.\ Rev.\ {\bf #1}}
\newcommand{\PRL}[1]{ Phys.\ Rev.\ Lett.\ {\bf #1}}

\newcommand{\be}{\begin{equation}}
\newcommand{\ee}{\end{equation}}
\newcommand{\ba}{\begin{eqnarray}}
\newcommand{\ea}{\end{eqnarray}}

\newcommand{\nn}{\nonumber}

\begin{document}
%
\begin{center}
\vspace{0.5cm}
\huge{\bf{Chiral symmetry and meson exchange }}

\vspace{0.4cm}
\huge{\bf{approach to hypernuclear decay}}
%

\end{center}
\vspace{.3cm}

\begin{center}
{\huge{E. Oset, D. Jido and J. E. Palomar}}
\end{center}

\begin{center}
{\small{\it Departamento de F\'{\i}sica Te\'orica e IFIC, Centro Mixto
Universidad de Valencia-CSIC,\\
46100 Burjassot (Valencia), Spain}}
\end{center}

\vspace{1cm}

\begin{abstract}
  We take an approach to the $\Lambda$ nonmesonic weak decay in nuclei based on the
  exchange of mesons under the guidelines of chiral Lagrangians. The one pion
   and one kaon exchange are considered,
  together with the exchange of two pions, either correlated, leading to an
  important scalar-isoscalar exchange
  ($\sigma$-like exchange), or uncorrelated
  (box diagrams).  A drastic reduction of the OPE results for the 
   $\Gamma_n/\Gamma_p$ ratio
   is obtained and the new results are compatible with all present experiments
   within errors. The absolute rates obtained for different nuclei are also in 
   fair agreement with experiment.
\end{abstract}
 [Key Word] $\Lambda$ weak decay,  $\Gamma_n/\Gamma_p$ ratio, chiral unitary
 theory.

\section{Introduction}

  The problem of the $\Gamma_n/\Gamma_p$ ratio is the most persistent and
  serious problem related to the nonmesonic decay of $\Lambda$ hypernuclei.
   The OPE
  model, using exclusively the parity conserving part of the weak $\Lambda$ 
  decay vertex $H_{\Lambda \pi N}$ leads to a   $\Gamma_n/\Gamma_p$ ratio of
  1/14 \cite{Oset:1990ey} in nuclear matter.  If in addition one includes the 
  parity violating term, which is
  less important than the parity conserving one for the nonmesonic decay, the
  ratio changes  to about 1/8 \cite{Dubach:1996dg,PRB97}.

     Experimentally one has results for $^5_\Lambda$He from 
  \cite{Szymanski:1991ik}  with a ratio 0.93$\pm$0.5 and for $^{12}_\Lambda$C with
  ratios $1.33^{+1.12}_{-0.81}$ \cite{Szymanski:1991ik}, $1.87^{+0.91}_{-1.59}$ 
  \cite{Noumi:1995yd} and 0.70$\pm$0.30, 0.52$\pm$0.16 \cite{montwill}. More recent
  results for $^{12}_\Lambda$C are still quoted as preliminary 
  \cite{outa,hashimoto} but also range in values around unity with large errors.
  
      The large discrepancy of the OPE predictions with the experimental data
      has stimulated much theoretical work. One line of progress has been the
      extension of the one meson exchange model including the exchange of
      $\rho,\eta,K,\omega,K^*$ in  \cite{Dubach:1996dg} and 
  \cite{PRB97}. The results obtained are somewhat contradictory since
  while in \cite{Dubach:1996dg} values for the $\Gamma_n/\Gamma_p$ ratio
 around 0.83 are quoted for $^{12}_\Lambda$C, the number quoted in 
 \cite{PRB97} is 0.07. Also, in \cite{PRB97} the same ratio
 is obtained for $^5_\Lambda$He and $^{12}_\Lambda$C while in 
 \cite{Dubach:1996dg} the value of the ratio in $^{12}_\Lambda$C is about twice
 larger than for $^5_\Lambda$He (see \cite{review} for a further discussion on
 this issue).
 
    Another line of progress has been the consideration of two pion exchange.
  An early attempt in \cite{shono} including N and $\Sigma$ intermediate states
  in a box diagram with two pions did not improve on the ratio and it made it
  actually slightly worse. However, in \cite{Shmatikov:1994up} the $\Delta$
  intermediate states were also considered leading to an increase of the 
   the $\Gamma_n/\Gamma_p$ ratio, although no numbers were given. A close line
   was followed in \cite{itonaga,Shmatikov:1994sp} where the exchange of two
   interacting pions through the $\sigma$ resonance was considered and found to
   lead also to improved results in the $\Gamma_n/\Gamma_p$ ratio. Although
   there are still some differences in the works and results of 
   \cite{itonaga,Shmatikov:1994sp} (see \cite{review} for details) they share
   the qualitative conclusion that the $\Gamma_n/\Gamma_p$ ratio increases when
   the $\sigma$ exchange is considered. In \cite{itonaga} the ratio goes from
   0.087 for only pion exchange to 0.14 when the correlated two pions in the 
   $\sigma$ channel (and also the $\rho$, which does not change much the ratio)
   are considered.
   
   Quark model inspired work leads to higher values for the 
   $\Gamma_n/\Gamma_p$  ratio from the contribution of short distances but the
   total rates are overpredicted \cite{Inoue:1998ep,oka,SIO00}.
   
     The situation is hence puzzling. Discrepancies between authors using the
    similar approach still persist, but in spite of that, there is a clear
    discrepancy between predictions of different models and present experimental
    results. 
    
      In the present talk I report on the recent work \cite{paper}, where in
    addition ot the one pion exchange we have considered kaon exchange and
    correlated as well as uncorrelated two pion exchange. The
    correlated two pion exchange has been done here following closely the steps
    of the recent work \cite{toki} where the two pions are allowed to interact
    using the Bethe-Salpeter equation and the chiral Lagrangians \cite{leu}.
    This chiral unitary approach to the pion pion scattering problem leads to
    good agreement with the $\pi \pi$ data in the scalar sector including the
    generation of a pole in the t-matrix corresponding to the $\sigma$ meson
    \cite{oller}.
    
     The results obtained here lead to ratios of $\Gamma_n/\Gamma_p$ of the 
     order of 0.4 and simultaneuosly one can obtain fair agreement for 
     the absolute rates
     of different nuclei. These relative high values obtained for 
     $\Gamma_n/\Gamma_p$ are compatible with all  present experiments within
     errors, if these errors are enlarged as suggested in  
     \cite{Ramos:1997ik} and \cite{gal95}.

\section{One Pion Exchange}
The  decay of the $\Lambda$ in nuclear matter was investigated with
the propagator approach which provides a unified picture of different
decay channels of the $\Lambda$ \cite{OS85}. The decay width of the $\Lambda$ is
calculated in  infinite nuclear matter, and is extended to finite
nuclei with the local density approximation. In this section we shall review the
calculation of the decay width of the $\Lambda$ in
nuclear matter using the one pion exchange approach.

First of all, we start with an effective $\pi \Lambda N$ weak interaction which is
written,
\begin{equation}
  {\cal L}_{\Lambda N \pi} = i G \mpi^2 \bar{\psi}_N
   [ A +  \gamma_5 B] \vec{\tau}\cdot \vec{\phi}_\pi \psi_\Lambda +
   {\rm h.c.} 
\end{equation}

\noindent where $\mpi$ denotes the pion mass, and 
$G$ is the weak coupling constant with
\begin{equation}
  G \mpi^2 = 2.211 \times 10^{-7}
\end{equation}

By assuming that the $\Lambda$  behaves as a $I=1/2, I_z=-1/2$ state in the 
isospin space, 
this effective interaction already implements the phenomenological
$\Delta I = 1/2$ rule, which is seen in the nonleptonic free decay of 
the $\Lambda$. 
The coupling constants $A$ and $B$ are determined by the parity
conserving and parity violating amplitudes of the 
nonleptonic $\Lambda$ decay, respectively:
\begin{equation}
  A = 1.06, \hspace{1cm} {B \over 2M_N} \mpi = - 0.527
\end{equation}
with $M_N$ the nucleon mass.
The $\pi NN$  vertex with strong interaction is given by the following
effective Lagrangian:
\begin{equation}
\label{lambNpi}
   {\cal L}_{\pi NN}^S =-\frac{D+F}{2f_\pi} \bar{\psi}_N \gamma^{\mu}\gamma_5
    \vec{\tau}\cdot \partial_{\mu}\vec{\phi_\pi} \psi_N
\end{equation}
\noindent with $D+F=1.26$ and $D-F=0.33$.

\begin{figure}[ht]
\centerline{\includegraphics[width=0.51\textwidth]{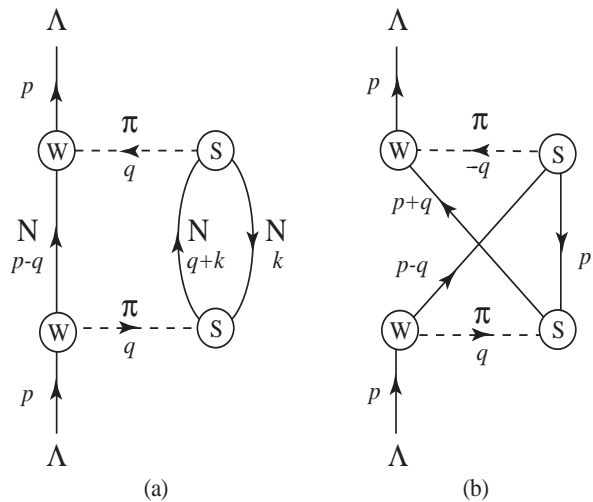}}
\caption{Lowest order of self-energy of $\Lambda$. The nonmesonic width
 comes from the imaginary part when the intermediate states cut by a horizontal
  line are placed on shell.}
\label{fig2}
\end{figure}

In order to evaluate the $\Lambda$ decay width $\Gamma$, in a nuclear
medium due to a certain $\Lambda N \rightarrow NN$ transition
amplitude, we start with the
calculation of the self-energy in the medium,  
$\Sigma$, shown in fig.\ref{fig2}, 
and then we take its imaginary part:
\begin{equation}
  \Gamma = - 2\ {\rm  Im }\ \Sigma
\end{equation}

In addition we take into account the $ph$ and $\Delta h$ excitations to all orders
in the sense of the random phase approximation (RPA) as done in \cite{OS85}.
\begin{figure}[ht]
\centerline{\includegraphics[width=1.0\textwidth]{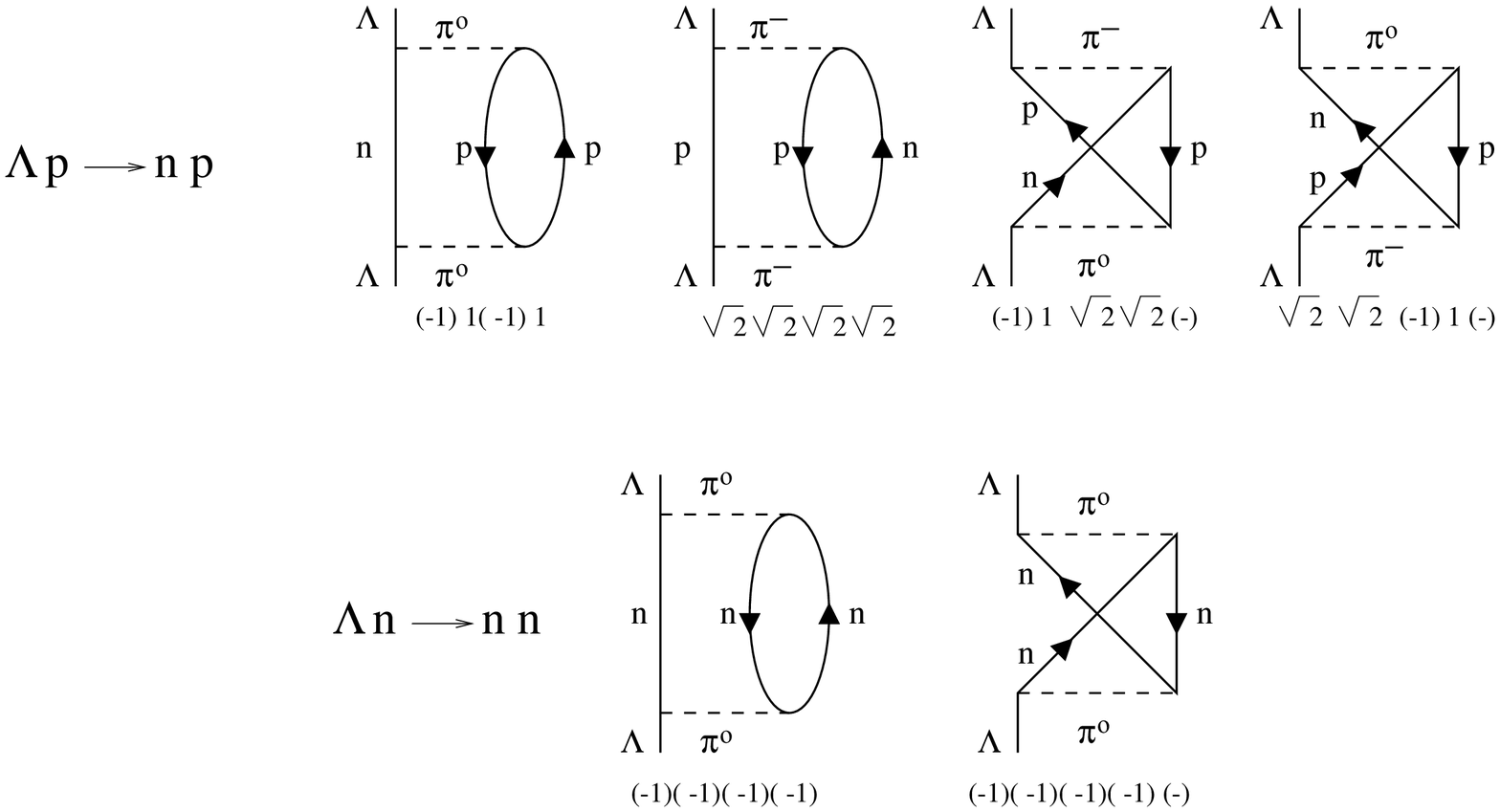}}
\caption{The isospin factors of the direct and exchange terms induced
by proton and neutron.}
\label{fig3}  
\end{figure}

On the other hand, short range correlations are also introduced following
\cite{OS85} and they modulate the $\Lambda N
\rightarrow NN$ transition amplitude. In terms of the Landau-Migdal parameter,
our value for $g'$ of the strong spin-isospin interaction has a strength of 0.7
at $\vec{q}=\vec{0}$ and a $\vec{q}$ dependence as given in~\cite{OS85}. 

In fig.~\ref{fig3} we can see the direct and exchange diagrams which contribute to the
$\Lambda$ decay induced by protons or neutrons. The representation is useful to see
the effects of the isospin in the $\Gamma_n/\Gamma_p$ rates.  First let us note
that the momentum of the pion in the upper pion of the exchange diagram is
 $-\vec{q}$ going to the left (neglecting Fermi motion), while in the direct
 diagram the momentum is $\vec{q}$.  This has as a consequence that the relative
 sign of the parity conserving versus parity violating terms in the upper pion
 exchange is opposite for the exchange diagram than for the direct one.
  As a consequence a
 simple counting of the rates is possible simply taking into account the isospin
 coupling of the vertices ($\sqrt{2}$ for charged pions, $1$ for the $\pi^0 pp$
 veçrtex and $-1$ for the $\pi^0 nn$ vertex). 
   We find the ratios for the parity conserving and parity violating parts:
   
   Parity conserving:
\begin{equation} \frac{\Gamma_{n}}{\Gamma_{p}}=\frac{1-\frac{1}{2}}{1+4+2\
\frac{1}{2}+
2\ \frac{1}{2}}=\frac{1}{14}
\end{equation}

Parity violating:
\begin{equation} \frac{\Gamma_{n}}{\Gamma_{p}}=\frac{1+\frac{1}{2}}{1+4-2\
\frac{1}{2}-
2\ \frac{1}{2}}=\frac{1}{2}
\end{equation}
   
     And given the weight of the parity conserving and parity violating parts
  one finds finally a ratio $\Gamma_n/\Gamma_p$  around 1/8.
     The ratio is too large  and the absolute
   rates, of about $\Gamma= 2 \Gamma_\Lambda$ are too large 
   compared with experiment.

\section{Kaon Exchange}
The non-mesonic decay of $\Lambda$ with one $K$ exchange takes place
through the diagram shown in fig.~\ref{fig2} substituting the pions by kaons in
the figure.
The inclusion of the $K$ exchange is  straightforward in the meson
propagator approach, once the $KNN$ weak vertex is fixed.

The strong $K\Lambda N$ vertex is given by:
\begin{equation}
   {\cal L}_{K\Lambda N}^S =  f_{KN\Lambda} \bar{\psi}
   \gamma^{\mu}\gamma_5 \partial_{\mu}\phi_{K} \psi_\Lambda + {\rm h.c.} 
\end{equation}
\noindent which 
is estimated with the $SU(3)$
flavor symmetry:

\begin{equation}
    f_{KN\Lambda} = {D+3F \over 2 \sqrt{3} f_\pi}
\end{equation}
Note that there is a different sign in
$f_{K\Lambda N}$ with respect to the $pp\pi^0$ vertex of eq.~(\ref{lambNpi}).

The weak vertex of $NNK$ may be written as
\begin{eqnarray}
\label{abk}
   {\cal L} &=& i G\mpi^2 \left[ \bar{\psi}_p (A^{K^0,p} +
   \gamma_5  B^{K^0,p} )\phi_{K^0}^\dagger \psi_p  \right. \nonumber \\
   && + \bar{\psi}_p (A^{K^-,p} + \gamma_5
   B^{K^-,p})\phi_{K^-}^\dagger \psi_n  \\
   && + \left. \bar{\psi}_n (A^{K^0,n} + \gamma_5
   B^{K^0,n})\phi_{K^0}^\dagger \psi_n   \right] \ + {\rm h.c.} \nonumber 
\end{eqnarray}

Equation~(\ref{abk}) shows the parity conserving (B coefficients) and parity violating
 terms (A coefficients) for the case of the kaon weak coupling. The parity
 violating terms can be deduced following \cite{Dubach:1996dg} by means of
 current algebra arguments assuming that they behave like the sixth component
 of the SU(3) generators or equivalently using appropriate chiral Lagrangians.
 On the other hand the parity conserving part does not follow this symmetry and
 some models have to be done.  We have used the results of the pole model used
 in \cite{Dubach:1996dg}.  As we shall see, the kaon exchange, through
 interference with the pions, leads both to higher $\Gamma_n/\Gamma_p$ ratios and
 also smaller total rates.

\section{Two-pion exchange}

Another kind of diagrams that have been traditionally studied are those
corresponding to two-pion exchange. We will divide the study of these diagrams
into two categories: correlated two-pion exchange and uncorrelated two-pion
exchange. In the case of correlated exchange we will only consider the scalar-
isoscalar channel, where the $\sigma$ meson appears. The vector channel is
neglected since the $\rho$ contribution has been seen to be not too 
relevant~\cite{Shmatikov:1994sp}. We will see that
the scalar-isoscalar channel is also the relevant one in the case of 
uncorrelated two-pion exchange.
 The effect of heavier scalar mesons (such as the $f_{0}, a_{0}$) is also 
 found negligible in ~\cite{Shmatikov:1994sp} and is neglected here.

\subsection{Correlated two-pion exchange}

Some works on this topic have been done~\cite{itonaga,Shmatikov:1994sp,iton2},
 incorporating
the $\sigma$ meson as an explicit degree of freedom. There it is found that,
working with reasonable values for the mass, width and $\sigma \pi \pi$
coupling, the role of this $"2\pi /\sigma"$
exchange is relevant in the non-mesonic decay problem.

A less phenomenological treatment of the sigma meson is provided by the Chiral
Unitary Approach~\cite{oller,IAM,N/D}. In~\cite{oller} it was found that the
$\sigma$ meson is dynamically generated by the
in-flight two pion interaction when summing up the s-wave $t-$matrix  of the
$\pi \pi$ scattering to all orders using the Bethe-Salpeter equation. The former
 picture of the $\sigma$ meson was used to describe its role in the $NN$
interaction in ref.~\cite{toki}, finding a moderate attraction beyond $r=0.9$ 
fm and a repulsion at shorter distances, in contrast with the all attraction of 
  the conventional $\sigma$ exchange. We will follow an analogous model to the one of
the aforementioned reference.

\begin{figure}[ht]
\centerline{\includegraphics[width=0.9\textwidth]{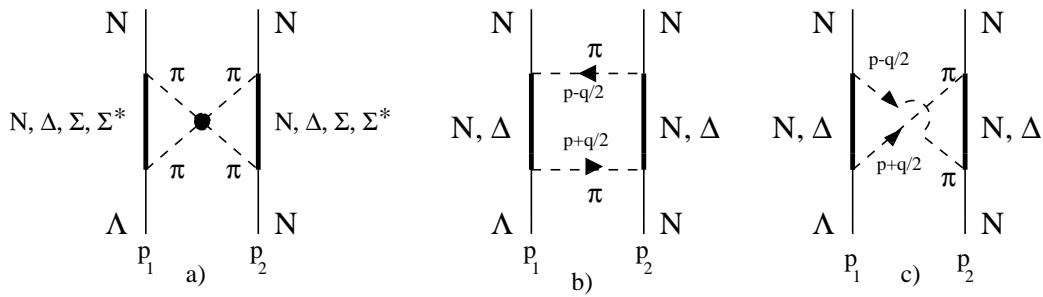}}
\caption{Diagrams corresponding to two-pion exchange: a) correlated exchange; b)
 uncorrelated exchange: direct diagram; c) uncorrelated exchange: crossed
 diagram.}
\label{Ndss}
\end{figure}

The diagrams corresponding to the correlated exchange are those of fig.
\ref{Ndss}a). In the weak vertex we will only consider the parity conserving
term of the lagrangians since it is the relevant one when dealing with loops.
This simplifies the problem because, as
 the parity conserving part (proportional to $\vec{\sigma}
 \vec{q}$, where $\vec{q}$ is the momentum of the pion) has the same structure
 as the $\pi NN$ interaction, the results obtained in ref.~\cite{toki}  are also
 applicable here.  One is then allowed to take the expression of the potential coming
from the diagrams with $N$ and $\Delta$ as intermediate states from that
reference, with the only
difference of a multiplicative factor $\mathcal{R}$ that reflects the replacement of one strong
 $\pi NN$ vertex by the weak $\Lambda \pi N$. The potential is then given
 by~\cite{toki}:
 
 \begin{equation}
 \label{tt}
 t_{\Lambda NNN}(q)=\mathcal{R} \ \widetilde{V}^{2}(q) \frac{6}{f^{2}}
\frac{\vec{q}^{\ 2}+\frac{m_{\pi}^{2}}{2}}{1-G(-\vec{q}^{\ 2})\frac{m_{\pi}^{2}}{2}{f^{2}}}
\end{equation}

\noindent where $G(s)$ is the loop function with two pion propagators, and the
vertex function 
$\widetilde{V}(q)$ and $\mathcal{R}$ are given by:

\ba
\label{rv}
\mathcal{R} &=& \frac{G \mu P}{\frac{D+F}{2f_\pi}}\nn\\
\widetilde{V}(q) &=& \widetilde{V}_{N}(q)+\widetilde{V}_{\Delta}(q)
\ea

So far we have been studying diagrams in which the $\Lambda$ baryon
appears in the weak vertices. However, as we can see in figure~\ref{Ndss}a), there
 are also diagrams with a strange intermediate baryon ($\Sigma, \Sigma^{*})$
 strongly coupled to the $\Lambda$ and the weak vertex in the upper side of the 
 diagram. We have evaluated them using SU(3) symmetry
 arguments and found an approximate cancellation between the 
 $\Sigma$ and $\Sigma^{*}$ contributions. 
 
 As one can see in eq.~(\ref{tt}) the sign of the $\sigma$ potential in momentum space
 is positive.  This is in contrast with any evaluation of the scalar-isoscalar
 potential taking only the exchange of a $\sigma$ particle.  Indeed one obtains 
 in that case the vertex
 squared times the $\sigma$ propagator, and the latter is always negative for
 the space like situation which one has here.  Thus, the chiral approach to  
 $\sigma$ exchange leads to an opposite sign than the ordinary $\sigma$ exchange
  contribution.

\subsection{Uncorrelated two-pion exchange}
The other set of processes that we have to study when considering the two-pion
exchange is the one in which the exchanged pions only interact with baryonic
legs and not with other pions (uncorrelated exchange). The corresponding Feynman
diagrams are depicted in figs.~\ref{Ndss}b) and \ref{Ndss}c).


We do not include the diagrams with an intermediate $\Sigma$  and
$\Sigma^{*}$,
 because one expects a similar cancellation to the one found in the correlated
 exchange, nor the diagrams with two nucleon propagators in diagram a), which
 correspond to final state interaction and are included in the correlations. 
  We also neglect the spin dependent term, which is found to be negligible.

\section{Results}

We show the results for $^{12}_{\Lambda}$C separating the different
contributions. In table~\ref{ope} we show the results obtained with only one
pion, one kaon or two pion  exchange. In addition we write there the
contributions when the kaon and the two pion cotributions are added coherently.


\begin{table}[ht]
\begin{center}
\begin{tabular}{|c|ccccc|}
\hline 
 \ \  & $\pi $ & $ K $ & $ 2\pi$ & $\pi +K$ & $\pi+K+2\pi$ \\
\hline
 $\Gamma_{p}/\Gamma_{free}$ & 0.9557 & 0.2527 & 0.1905 & 0.5215 & \ \ \ 0.5714\\
 $\Gamma_{n}/\Gamma_{free}$ & 0.1194 & 0.1180 & 0.0903 & 0.2728 & \ \ \ 0.3078\\ 
\hline
\end{tabular}
\label{ope}
\caption{$\pi$, K, $2\pi$ contributions together with the combinations of
$\pi+K$, $\pi+K+2\pi$ to the proton- and neutron-induced decay of
$^{12}_{\Lambda}$C.}
\end{center}
\end{table}

While $K$ or $2\pi$ contributions by themselves are small compared to
$\Gamma_{p}$ from OPE, the interference effects with the OPE contribution are
large. We can see that the introduction of the kaon exchange reduces the proton
rate by about a factor two and increases the neutron rate also by about a factor
two, thus increasing the ratio in about a factor four and reducing the total
rate. The additional effects of the two pion contribution are small in both
rates as a consequence of some cancellations.
It is worth recalling, as we mentioned above, that the $\sigma$ and
uncorrelated $2\pi$ contributions have different signs and there are large
cancellations between them at the relevant momentum $q\sim 420$ MeV/c. Let us
stress once more that we obtain a sign for the $\sigma$ exchange here which is
opposite to the conventional one.
Should we have the $\sigma$ contribution with opposite sign to ours and about
the same strength, the combination of $\sigma$ and uncorrelated two-pion
exchange would give a contribution for the $2\pi$ part alone about 6 times bigger
than here, and this would render the total rates unacceptably large in spite of
the interference terms, which are only multiplied by a factor 2.5.

In table~\ref{final} we present the results for $\Gamma_p$, $\Gamma_n$ and the
$\Gamma_{n}/\Gamma_{p}$ ratio for different nuclei. We find that the total rates from the $1p1h$ channel 
 go from $\Gamma/\Gamma_{free}=0.88$ to
1.48 in $^{208}_{\Lambda}$Pb and the ratios $\Gamma_{n}/\Gamma_{p}$ are all of
them of about $\Gamma_{n}/\Gamma_{p}\sim 0.54$.

\begin{table}[ht]
\begin{center}
\begin{tabular}{cccccc}
\hline 
 Nucleus & $\Gamma_p/\Gamma_{free}$ & $\Gamma_n/\Gamma_{free}$ & $(\Gamma_{p}+
 \Gamma_{n})/\Gamma_{free}$ & $\Gamma_{n}/\Gamma_{p}$ &
 $\Gamma_{tot}/\Gamma_{free}$\\ 
\hline
 $^{12}_{\Lambda}$C & 0.5714 & 0.3078 & 0.8792 & 0.54 & 1.3992\\
 $^{28}_{\Lambda}$Si & 0.7562 & 0.4080 & 1.1642 & 0.54 &1.5342\\
 $^{40}_{\Lambda}$Ca & 0.7866 & 0.4245 & 1.2111 & 0.54 & 1.5411\\
 $^{56}_{\Lambda}$Fe & 0.8554 & 0.4620 & 1.3174 & 0.54 & 1.6174\\
 $^{89}_{\Lambda}$Y & 0.8908 & 0.4813 & 1.3721 &  0.54 & 1.6721\\
 $^{139}_{\Lambda}$La & 0.8702 & 0.4697 & 1.3399 &  0.54& 1.6399\\
 $^{208}_{\Lambda}$Pb & 0.9640 & 0.5195 &  1.4835 &  0.54 & 1.7835\\
\hline
\end{tabular}
\label{final}
\caption{Decay rates and the $\Gamma_{n}/\Gamma_{p}$ ratio for different
hypernuclei. In $\Gamma_{tot}$ we have included the contributions from the
mesonic decay and the $2p2h$ channel.}
\end{center}
\end{table}

If we want to compare these results with experimental data we should still add
the mesonic contribution and the $2p2h$ induced one. For the mesonic
contribution we take the results from~\cite{nieves} which agree well with
experiment in the measured cases. The mesonic rates are only relevant for the
lighter nuclei. We take $\Gamma_{M}/\Gamma_{free}=0.25$ for $^{12}_{\Lambda}$C,
0.07 for $^{26}_{\Lambda}$Si and 0.03 for $^{40}_{\Lambda}$Ca and neglect this
contribution for heavier nuclei. The $2p2h$ induced contribution calculated
in~\cite{ROS95} is 0.27 for $^{12}_{\Lambda}$C and 0.30 for the rest
of the nuclei. With these results we compute the total rates which we show in
table 2. The present status of the lifetime measurements can be found in
table~\ref{park}. We can see that our total rates are about $15\%$ bigger than
the experimental numbers in the best measured nuclei. In heavy nuclei the
experimental errors are larger and our results are compatible with the
experiment.

\begin{table}[ht]
\begin{center}
\begin{tabular}{ccc}
\hline 
 Nucleus & $\Gamma/\Gamma_{free}$ & Experiment \\ 
\hline
  
   $^{11}_{\Lambda}$B & $1.37\pm 0.17$\cite{16s,17s} & $(K^{-},\pi^{-})$ \\
   & $1.25\pm 0.08$\cite{hpark} & $(K^{+},\pi^{+})$ \\
   $^{12}_{\Lambda}$C & $1.25\pm 0.19$\cite{16s,17s} & $(K^{-},\pi^{-})$ \\
   & $1.14\pm 0.08$\cite{hpark} & $(K^{+},\pi^{+})$ \\
   $^{28}_{\Lambda}$Si & $1.28\pm 0.08$\cite{hpark} & $(K^{+},\pi^{+})$ \\
   $_{\Lambda}$Fe & $1.22\pm 0.08$\cite{hpark} & $(K^{+},\pi^{+})$ \\
   $\bar{p}+^{209}$Bi & $1.1^{+1.1}_{-0.4}$\cite{19s} & Delayed fission \\
    & $1.5\pm 0.3\pm 0.5$\cite{20s} & Delayed fission \\
   $p+^{209}$Bi & $1.8\pm 0.1\pm 0.3$\cite{Kulessa:1998kg} & Delayed fission \\
\hline
\end{tabular}
\label{park}
\caption{Experimental values of the total width for different nuclei. The value
for $_{\Lambda}$Fe represents for the average lifetime of $^{55}_{\Lambda}$Mn, 
$^{55}_{\Lambda}$Fe and $^{56}_{\Lambda}$Fe.}
\end{center}
\end{table}
\normalsize

As for the $\Gamma_{n}/\Gamma_{p}$ ratios it looks like our results are still
smaller than the experimental ones. However, one word of caution is necessary
here. The experimental analyses were done neglecting the $2p2h$ induced channel,
but it was observed in~\cite{ROS95} that the inclusion of this mechanism in the
analysis of the data lead to different values of $\Gamma_{n}/\Gamma_{p}$. A
formula was given in this reference to correct the results of the old analysis
due to the consideration of this induced mechanism, but it assumed that all
particles were detected. The formula was corrected in \cite{gal95} assuming
 that the
slow particles (with energies smaller than about 40 MeV) are not detected.
Detailed calculations of the spectra of protons and neutrons from the nonmesonic
decay were done in~\cite{Ramos:1997ik} but assuming a ratio of $1p1h$ to $2p2h$ induced
strength given by the OPE model alone, which as shown here overcounts the $1p1h$
strength. In view of this we just take the formula of~\cite{gal95} and use it to
recalculate the experimental bands. The present bands we have are:
$1.33^{+1.12}_{-0.81}$\cite{Szymanski:1991ik}, $1.87^{+0.91}_{-1.59}$
\cite{Noumi:1995yd}, $0.70\pm
0.30$\cite{montwill}, $0.52\pm 0.16$\cite{montwill}. The lower bounds 
are 0.52, 0.29, 0.4, 0.36 respectively.
Our ratio 0.54 falls within all the error bands, but close to the lower
boundary. However, if we use the formula of~\cite{gal95} assuming $\Gamma_{2p2h}/\Gamma_{nm}$
of the order of 0.3 one reduces the lower bounds to values 0.2, 0.14, 0.1, 0.01
and the value obtained by us is well within present experimental ranges.

\section{Conclusions}

We have evaluated the nonmesonic proton and neutron induced $\Lambda$ decay
rates in nuclei, by including one pion, one kaon, $\sigma$ 
and uncorrelated two pion exchange. We found that the contribution of $K$
exchange was essential to reduce the total decay rate from the OPE results and
simultaneously increase the value of the $\Gamma_{n}/\Gamma_{p}$ ratio from
values around 0.12 for the OPE to values around 0.54. We also included the
$\sigma$ and uncorrelated two pion exchange and we found some cancellations
between them, such that the total contribution of the two pion exchange to the
total rate and the $\Gamma_{n}/\Gamma_{p}$ ratio was small. However, in this
result it was very important that the contribution of our correlated
scalar-isoscalar two-pion exchange had
opposite sign to the conventional contributions taking only the exchange of a  
$\sigma$ particle. This change of sign was due to the presence
of the Adler zero in the scalar-isoscalar $\pi \pi$ interaction which makes the
amplitude change sign below $s=m_{\pi}^{2}/2$ which is the case here, where we
have $s$ negative.

The total rates obtained are fair, about $15\%$ larger than experiment as an
average. The ratios $\Gamma_{n}/\Gamma_{p}$ are considerably improved with
respect to the OPE ones. We have also seen that, once the present experimental
data are corrected to account for the $2p2h$ channel the value of 0.54 obtained
here for the $\Gamma_{n}/\Gamma_{p}$ ratio is well within the present
experimental boundaries.

\subsection*{Acknowledgements}
We wish to acknowledge multiple and useful discussions with A. Ramos and A.
Parre\~no. Useful information from A. Gal and M. Oka is much appreciated. 
 We would also like to acknowledge financial
support from the DGICYT under contracts PB96-0753, PB98-1247 and
AEN97-1693, from the Generalitat de Catalunya under grant
SGR98-11
and from the EU TMR network Eurodaphne, contract no.
ERBFMRX-CT98-0169. J. E. Palomar and D. Jido acknowledge support from Ministerio
de Ciencia y Tecnolog\'{\i}a.




\begin{thebibliography}{200}
\bibitem{Oset:1990ey}
E.~Oset, P.~Fernandez de Cordoba, L.~L.~Salcedo and R.~Brockmann,
Phys.\ Rept.\  {\bf 188}, 79 (1990).


\bibitem{Dubach:1996dg}
J.~F.~Dubach, G.~B.~Feldman and B.~R.~Holstein,
Annals Phys.\  {\bf 249}, 146 (1996)
[nucl-th/9606003].

\bibitem{PRB97}
A.~Parreno, A.~Ramos and C.~Bennhold,
Phys.\ Rev.\  {\bf C56}, 339 (1997) 
[nucl-th/9611030] and private communication of recent results.


\bibitem{Szymanski:1991ik}
J.~J.~Szymanski {\it et al.},
Phys.\ Rev.\  {\bf C43}, 849 (1991).

\bibitem{Noumi:1995yd}
H.~Noumi {\it et al.},
Phys.\ Rev.\  {\bf C52}, 2936 (1995).



\bibitem{montwill}
A. Montwill et al., Nucl. Phys. {\bf A234} (1974) 413.

\bibitem{outa} H.Outa, Nucl. Phys. A, Conf. Proceedings,  A670(2000)281c

\bibitem{hashimoto} O.Hashimoto, private communication.

\bibitem{review} E. Oset and A. Ramos, Prog. Part. Nucl. Phys. 41 (1998) 191.
  
\bibitem{shono} H. Band\={o}, Y. Shono and H. Takaki,
Int. J.  Mod. Phys. {\bf A3} (1988) 1581.

\bibitem{Shmatikov:1994up}
M.~Shmatikov,
Phys.\ Lett.\  {\bf B322}, 311 (1994).

\bibitem{itonaga}
K. Itonaga, T. Ueda, and T. Motoba, Nucl. Phys. {\bf A585}
(1995) 331c; {\it ibid.}, {\it Weak and Electromagnetic
Interactions in
Nuclei}, edited by H.
Ejiri, T. Kishimoto and T. Sato (World Scientific, Singapore,
1995) 546.

\bibitem{Shmatikov:1994sp}
M.~Shmatikov,
Nucl.\ Phys.\  {\bf A580}, 538 (1994).



\bibitem{Inoue:1998ep}
T.~Inoue, M.~Oka, T.~Motoba and K.~Itonaga,
Nucl.\ Phys.\  {\bf A633}, 312 (1998)

\bibitem{oka} M. Oka, Talk given at the International Symposium on Nuclear 
Electro-Weak Spectroscopy (NEWS 99) for Symmetries and Electro-Weak
Nuclear-Processes, Osaka, Japan, 9-12 Mar 1999. 
e-Print Archive: nucl-th/9906041 

\bibitem{SIO00} K. Sasaki, T. Inoue, M. Oka, Nucl. Phys. A669, 331,
(2000), (E){\it ibid.} A678, 455, (2000).


\bibitem{paper} D. Jido, E. Oset and J.E. Palomar, IFIC preprint, University of
Valencia.

\bibitem{toki} E. Oset, H. Toki, M. Mizobe and T.T. Takahashi, Prog. Theor. Phys.
103 (2000) 351.

\bibitem{leu} J. Gasser and H. Leutwyler, \AN{158} (1984) 142; J. Gasser 
and H. Leutwyler, \NP{B250} (1985) 465, 517, 539; A. Pich, Rep. Prog.
 Phys. 58 (1995) 563; G. Ecker, Prog. Part. Nucl. Phys. 35 (1995) 1; U.G.
 Mei$\ss$ner, Rep. Prog. Phys. 56 (1993) 903.
 
\bibitem{oller} J.A. Oller and E. Oset, Nucl. Phys. A 620 (1997) 438; Erratum,
A652 (1999) 407.



\bibitem{Alberico:1991wg}
W.~M.~Alberico, A.~De Pace, M.~Ericson and A.~Molinari,
Phys.\ Lett.\  {\bf B256}, 134 (1991).


\bibitem{ROS95} A. Ramos, E. Oset and L.L. Salcedo, Phys. Rev. C {\bf
50}, 2314 (1994).


\bibitem{Ramos:1997ik}
A.~Ramos, M.~J.~Vicente-Vacas and E.~Oset,
Phys.\ Rev.\  {\bf C55}, 735 (1997)

\bibitem{gal95} A. Gal,
{\it Weak and Electromagnetic Interactions in
Nuclei}, edited by H.
Ejiri, T. Kishimoto and T. Sato (World Scientific, Singapore,
1995) 573.


\bibitem{OS85} E. Oset and L.L. Salcedo, Nucl. Phys. {\bf A443}, 704
               (1985).


\bibitem{iton2} K. Itonaga, T. Ueda and T. Motoba, \NP{A639} (1998) 329
\bibitem{IAM} J. A. Oller, E. Oset and J. R. Pel\'aez, \PRL{80} (1998), 3452
\bibitem{N/D} J. A. Oller and E. Oset, \PR{D60} (1999), 074023

\bibitem{Bhang:1998zp}
H.~Bhang {\it et al.},
Phys.\ Rev.\ Lett.\  {\bf 81}, 4321 (1998).

\bibitem{nieves} J. Nieves and E. Oset, Phys.\ Rev.\  {\bf C47}, 1478 (1993).
\bibitem{16s} J. J. Szymanski $et\ al.$, \PR{C43}, 849 (1991).
\bibitem{17s} R. Grace $et\ al.$, \PRL{55}, 1055 (1985).
\bibitem{hpark} H. Park $et\ al.$, \PR{C61}, 054004 (2000).
\bibitem{19s} J. P. Bocquet $et\ al.$, \PL{B192}, 312 (1987);{\bf 182}, 146 (1986).
\bibitem{20s} T. A. Armstrong $et\ al.$, \PR{C47}, 1957 (1993).
Phys. {\bf 43}, 856 (1986)].


\bibitem{Kulessa:1998kg}
P.~Kulessa {\it et al.},
Phys.\ Lett.\  {\bf B427}, 403 (1998).



\end{thebibliography}
\end{document}